\documentclass[
prl,
twocolumn,
showpacs,
amsmath,
floatfix,
superscriptaddress
]{revtex4}
\usepackage{graphicx}
\begin{document}
%
\title{Determination of the Parity of the Neutral Pion via the
       Four-Electron Decay}
%

\newcommand{\UAz}{University of Arizona, Tucson, Arizona 85721}
\newcommand{\UCLA}{University of California at Los Angeles, Los Angeles,
                    California 90095}
\newcommand{\Campinas}{Universidade Estadual de Campinas, Campinas,
                       Brazil 13083-970}
\newcommand{\EFI}{The Enrico Fermi Institute, The University of Chicago,
                  Chicago, Illinois 60637}
\newcommand{\UB}{University of Colorado, Boulder, Colorado 80309}
\newcommand{\ELM}{Elmhurst College, Elmhurst, Illinois 60126}
\newcommand{\FNAL}{Fermi National Accelerator Laboratory,
                   Batavia, Illinois 60510}
\newcommand{\Osaka}{Osaka University, Toyonaka, Osaka 560-0043 Japan}
\newcommand{\Rice}{Rice University, Houston, Texas 77005}
\newcommand{\SaoPaulo}{Universidade de S\~ao Paulo, S\~ao Paulo, Brazil 05315-970}
\newcommand{\UVa}{The Department of Physics and Institute of Nuclear and
                  Particle Physics, University of Virginia,
                  Charlottesville, Virginia 22901}
\newcommand{\UW}{University of Wisconsin, Madison, Wisconsin 53706}

\affiliation{\UAz}
\affiliation{\UCLA}
\affiliation{\Campinas}
\affiliation{\EFI}
\affiliation{\UB}
\affiliation{\ELM}
\affiliation{\FNAL}
\affiliation{\Osaka}
\affiliation{\Rice}
\affiliation{\SaoPaulo}
\affiliation{\UVa}
\affiliation{\UW}

\author{E.~Abouzaid}  \affiliation{\EFI}
\author{M.~Arenton}       \affiliation{\UVa}
\author{A.R.~Barker}      \altaffiliation[Deceased.]{ } \affiliation{\UB}
\author{L.~Bellantoni}    \affiliation{\FNAL}
\author{E.~Blucher}       \affiliation{\EFI}
\author{G.J.~Bock}        \affiliation{\FNAL}
\author{E.~Cheu}          \affiliation{\UAz}
\author{R.~Coleman}       \affiliation{\FNAL}
\author{M.D.~Corcoran}    \affiliation{\Rice}
\author{B.~Cox}           \affiliation{\UVa}
\author{A.R.~Erwin}       \affiliation{\UW}
\author{C.O.~Escobar}     \affiliation{\Campinas}  
\author{A.~Glazov}        \affiliation{\EFI}
\author{A.~Golossanov}    \affiliation{\UVa} 
\author{R.A.~Gomes}       \affiliation{\Campinas}
\author{P. Gouffon}       \affiliation{\SaoPaulo}
\author{Y.B.~Hsiung}      \affiliation{\FNAL}
\author{D.A.~Jensen}      \affiliation{\FNAL}
\author{R.~Kessler}       \affiliation{\EFI}
\author{K.~Kotera}  \affiliation{\Osaka}
\author{A.~Ledovskoy}     \affiliation{\UVa}
\author{P.L.~McBride}     \affiliation{\FNAL}

\author{E.~Monnier}
   \altaffiliation[Permanent address ]{C.P.P. Marseille/C.N.R.S., France}
   \affiliation{\EFI}  

\author{H.~Nguyen}       \affiliation{\FNAL}
\author{R.~Niclasen}     \affiliation{\UB}
\author{D.G.~Phillips~II} \affiliation{\UVa}
\author{E.J.~Ramberg}    \affiliation{\FNAL}
\author{R.E.~Ray}        \affiliation{\FNAL}
\author{M.~Ronquest}     \affiliation{\UVa}
\author{E.~Santos}       \affiliation{\SaoPaulo}
\author{W.~Slater}       \affiliation{\UCLA}
\author{D.~Smith}        \affiliation{\UVa}
\author{N.~Solomey}      \affiliation{\EFI}
\author{E.C.~Swallow}    \affiliation{\EFI}\affiliation{\ELM}
\author{P.A.~Toale}       \altaffiliation{Current address: Pennsylvania State University, University Park, Pennsylvania 16802}  \affiliation{\UB}
\author{R.~Tschirhart}   \affiliation{\FNAL}
\author{Y.W.~Wah}        \affiliation{\EFI}
\author{J.~Wang}         \affiliation{\UAz}
\author{H.B.~White}      \affiliation{\FNAL}
\author{J.~Whitmore}     \affiliation{\FNAL}
\author{M.~J.~Wilking}      \affiliation{\UB}
\author{B.~Winstein}     \affiliation{\EFI}
\author{R.~Winston}      \affiliation{\EFI}
\author{E.T.~Worcester}  \affiliation{\EFI}
\author{T.~Yamanaka}     \affiliation{\Osaka}
\author{E.~D.~Zimmerman} \affiliation{\UB}
\author{R.F.~Zukanovich} \affiliation{\SaoPaulo}

\collaboration{The KTeV Collaboration}
\thanks{Correspondence should be addressed to P.~A.~Toale or E.~D.~Zimmerman. Electronic addresses: {\tt toale@phys.psu.edu}, {\tt edz@colorado.edu}}\noaffiliation
\date{February 14, 2008}
\begin{abstract}
 We present a new determination of the parity of
the neutral pion via the double Dalitz decay $\pieeee$.  Our sample,
which consists of 30~511 candidate decays, was collected from
$\klpipipi$ decays in flight at the KTeV-E799 experiment at Fermi
National Accelerator Laboratory.  We confirm the negative $\pi^0$
parity, and place a limit on scalar contributions to the $\pieeee$
decay amplitude of less than 3.3\% assuming $\oCPT$ conservation.  The
$\piggvert$ form factor is well described by a momentum-dependent
model with a slope parameter fit to the final state phase space
distribution.  Additionally, we have measured the branching ratio of
this mode to be $B(\pieeee) = (3.26 \pm 0.18)\times 10^{-5}$.
\end{abstract}
\maketitle
\begin{figure}
\includegraphics[height=1.5in]{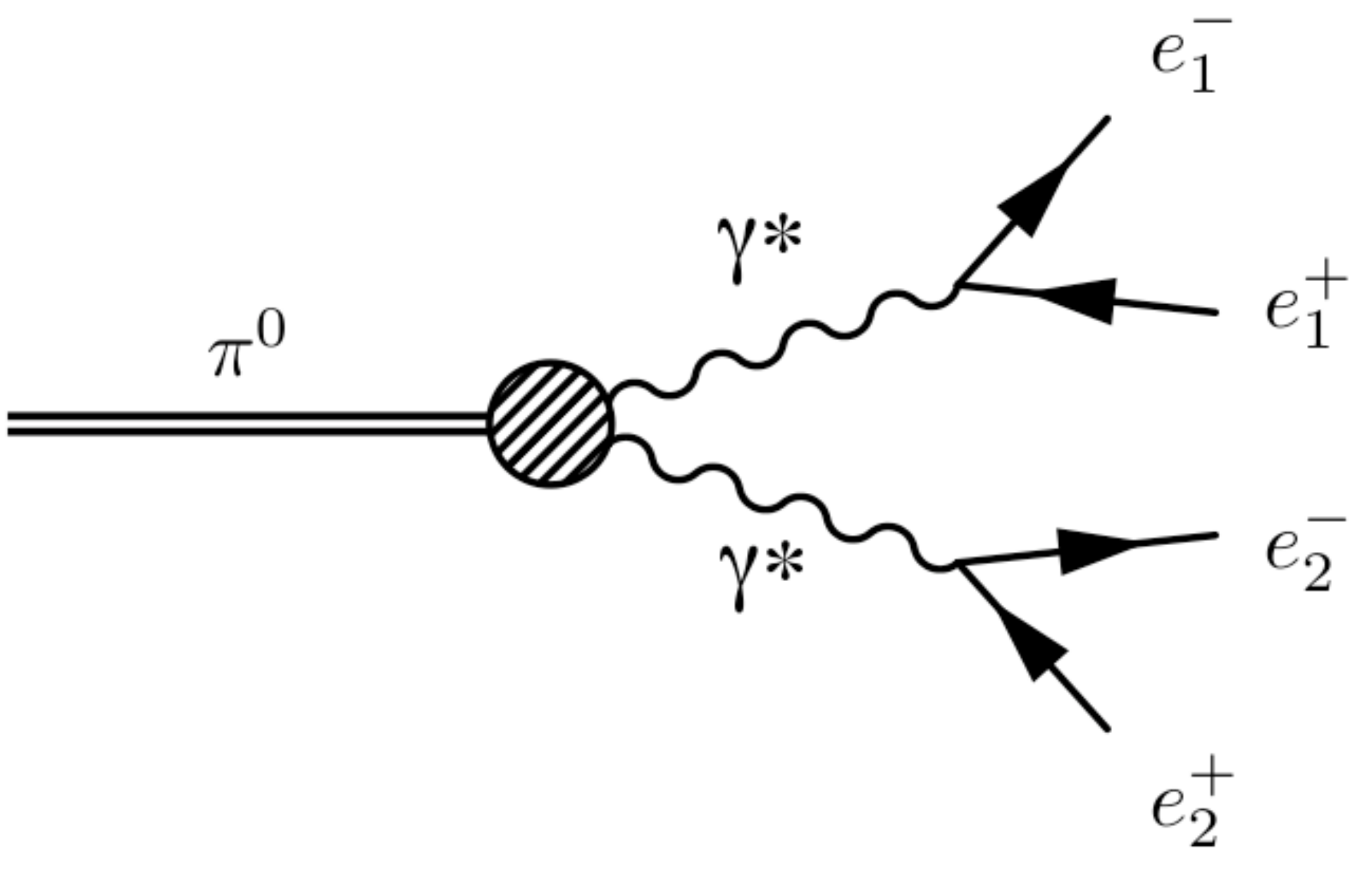}
\caption{\label{ddfeyn} Lowest order Feynman diagram for $\pieeee$.
  The direct contribution is shown; a second diagram exists with
  $e_1^+$ and $e_2^+$ exchanged.}
\end{figure}

%
%
The parity of the neutral pion has been determined indirectly by
studying negative pions captured on
deuterium~\cite{panofsky,chinowsky}.  The observed reactions imply
that the $\pi^-$ is a pseudoscalar and that the parities of the
$\pi^-$ and the $\pi^0$ are the same. It has long been known that the
decay $\pigg$ in principle offers a direct means of determining the
$\pizero$ parity through the polarizations of the
photons~\cite{yang,bernstein}. Given that there are no available
methods for measuring the polarization of a high-energy photon, this
measurement has never been performed. However, it was soon noted that
the double Dalitz decay $\pieeee$, which proceeds through an
intermediate state with two virtual photons (see Fig.\ \ref{ddfeyn}),
is sensitive to the parity of the pion since the plane of a Dalitz
pair is correlated with the polarization of the virtual
photon~\cite{kroll,dalitz}.  This process was studied in a 1962
hydrogen bubble chamber experiment using stopping negative pion
capture ($\pi^- p \rightarrow n \pi^0$).  That group observed 206
$\pieeee$ events and reported that the observed distribution of the
$e^+e^-$ planes was consistent with a pseudoscalar pion and disfavored
a scalar pion at the level of 3.6 standard deviations~\cite{samios2};
this experiment also produced a measurement of the branching ratio of
this decay, which remains the most precise result to date.

Using a sample of more than 30~000 $\pieeee$ decays, we report new
precise measurements of the properties of this decay.  Our modeling of
the decay includes for the first time a proper treatment of the
exchange contribution to the matrix element, and consideration of full
$\mathcal{O}(\alpha^2)$ radiative corrections.  With these advances,
we have tested for a scalar contribution in the $\piggvert$ coupling
with a sensitivity of a few percent.  We have also measured for the
first time the momentum dependence of the form factor in this decay
mode.  In addition, we present a new measurement of the $\pieeee$ branching
ratio, taking into account radiative effects.

The most general interaction Lagrangian for the $\pi^0\rightarrow\gamma^*\gamma^*$ transition
can be written \cite{barker}:
\begin{equation}
{\cal L} \propto C_{\mu\nu\rho\sigma}F^{\mu\nu}F^{\rho\sigma}\Phi
\end{equation}
where $F^{\mu\nu}$ and $F^{\rho\sigma}$ are the photon fields, $\Phi$
is the pion field, and the coupling has the form
\begin{equation}\begin{split}
C_{\mu\nu\rho\sigma} \propto & f(x_1, x_2)
      [ \cos\zeta\couplep \\*
    & \quad\quad\quad + \sin\zeta\me^{\mi\delta}\couples].
\end{split}\end{equation}
The first term in $C_{\mu\nu\rho\sigma}$ is the expected pseudoscalar
coupling and the second term introduces a scalar coupling with a
mixing angle $\zeta$ and a phase difference $\delta$.  Nuclear parity
violation would introduce a nonzero $\zeta$, while $CPT$ violation
would cause the phase $\delta$ to be nonzero.  We assume the standard
parity-conserving form for the $\gamma^*\rightarrow e^+e^-$
conversion.

The form factor $f(x_1,x_2)$ is expressed in terms of the momentum
transfer of each of the virtual photons, or equivalently the invariant
masses of the two Dalitz pairs: $x_{1} \equiv
(m_{e^+_{1}e^-_{1}}/M_{\pi^0})^2; x_{2} \equiv
(m_{e^+_{2}e^-_{2}}/M_{\pi^0})^2$.  In calculating the phase space
variables for an individual event, there is an intrinsic ambiguity in
assigning each electron to a positron to form a Dalitz pair.  Our
analysis uses a matrix element model that includes the exchange
diagrams and therefore avoids the need to enforce a pairing choice.

The $\piggvert$ form factor has been studied previously in the decay
$\pieeg$~\cite{fonvieille,farzanpay,drees}, where the quantity of
interest has been the slope parameter $a$ of the first-order Taylor
expansion $f(x,0) = 1 + ax$, with $x \equiv
m_{e^+e^-}^2/M_{\pizero}^2$. Here we use a form factor parametrizaton
based on the model of D'Ambrosio, Isidori, and Portol\'{e}s
(DIP)~\cite{dip}, but with an additional constraint that ensures the
coupling vanishes at large momenta~\cite{toalethesis}. In terms of the
remaining free parameters, the form factor is:
\begin{equation}
f_{\text{DIP}}(x_1, x_2;\alpha) = \frac{1 - \mu (1+\alpha) (x_1+x_2)}
                                         {(1-\mu x_1)(1-\mu x_2)},
\end{equation}
where $\mu = M_{\pizero}^2/M_\rho^2 \approx 0.032$. In the limit of
small $x$, this coincides with the Taylor expansion provided
$a = - \mu\alpha$.

The parity properties of the decay can be extracted from the angle
$\phi$ between the planes of the two Dalitz pairs in
Fig.~\ref{ddfeyn}, where pair 1 is defined as having the smaller
invariant mass.  The distribution of this angle from the
dominant direct contribution has the form $d \Gamma / d \phi \sim 1 -
A \cos (2\phi) + B \sin (2\phi)$, where $A\approx 0.2 \cos (2\zeta)$
and $B\approx 0.2\sin(2\zeta)\cos\delta$. A pure pseudoscalar
coupling, therefore, would produce a negative $\cos (2\phi)$
dependence.

%
%
The $\pizero$ decays used in this analysis are the result of
fully-reconstructed $\klpipipi$ decays in flight collected by the
KTeV-E799 experiment at Fermilab.  The E799-II experiment and the KTeV
detector are described elsewhere \cite{ktev799,ktev_eps}.  This
analysis relies on two core systems of the KTeV detector: a drift
chamber-based charged particle spectrometer and a cesium iodide (CsI)
electromagnetic calorimeter.  Electrons are identified as charged
particles whose entire energy is deposited in the CsI, while photons
are reconstructed from electromagnetic showers in the CsI with no
associated charged tracks.

The signal mode, denoted by $\klpipipidd$ where $\pi^0_{\rm DD}$ refers to
$\pieeee$, has
a signature of four charged particles identified as electrons and
with a combined invariant mass consistent with the $\pizero$ mass, plus
four photons that are compatible with two additional $\pizero$'s. 
Furthermore, the eight-particle state has an invariant mass
consistent with the $\klong$ and total momentum vector
in the direction of the kaon line of flight.

The branching ratio measurement, which we describe here first, makes
use of a normalization mode in which two pions decay via $\pieeg$ and
the third $\pigg$. This ``double single-Dalitz'' mode, denoted
$\klpipidpid$ where $\pi^0_{\rm D}$ refers to $\pieeg$, has the same
final state particles as the signal mode and is again identified by
finding the proper combinations of particles to make three pions with
a total momentum consistent with the kaon. The similarity of these
modes allows cancellation of most detector-related systematic effects
in the branching ratio measurement, but also allows each mode to be a
background to the other.

Radiative corrections complicate
the definition of the Dalitz decays in general.  We define the signal
mode $\pieeee$ to be inclusive of radiative final states where the
squared ratio of the invariant mass of the four electrons to the
neutral pion mass $x_{4e}\equiv (M_{4e}/M_{\pi^0})^2$ is greater than
$0.9$, while events with $x_{4e}<0.9$ (approximately 6\% of the total
rate) are treated as $\pieeeeg$.  
For normalization, the decay $\pieeg$ is
understood to include all radiative final states, for consistency with
previous measurements of this decay \cite{schardt}.  
Radiative corrections in this analysis are taken from an analytic
calculation to order $\mathcal{O}(\alpha^2)$ ~\cite{barker}.  

Other final states of the $\klpipipi$ decay can become backgrounds to
either the signal or normalization mode if one or more photons convert
to an $e^+e^-$ pair in the detector material: $\klpipipid$ where one
of the five photons converts, or $\klpipipi \rightarrow 6\gamma$ where
two photons convert.  These modes again have the same final state as
the signal, but can be distinguished statistically since the
externally produced pairs tend to have smaller invariant masses than
those from internal conversions.  The most significant of these
backgrounds is $\klpipipid$ with one external conversion in
material. The photon must convert upstream of the first drift chamber
for the resulting tracks to be reconstructed. The material in
this region sums to $2.8\times 10^{-3}$ radiation lengths. With five
photons available, the probability of one converting is
$\units{1.08}{\%}$, close to the single-Dalitz branching ratio. The
distinguishing characteristic of these events is the small value of
the $e^+e^-$ invariant mass, or similarly, the small value of the
opening angle of the pair.  Requiring a track separation at the first
drift chamber of greater than $\units{2}{\mm}$ removes
$\units{99.74}{\%}$ of the remaining simulated background while preserving
$\units{74.3}{\%}$ of signal and $\units{72.7}{\%}$ of normalization events.

\begin{figure}
\includegraphics[width=3.5in]{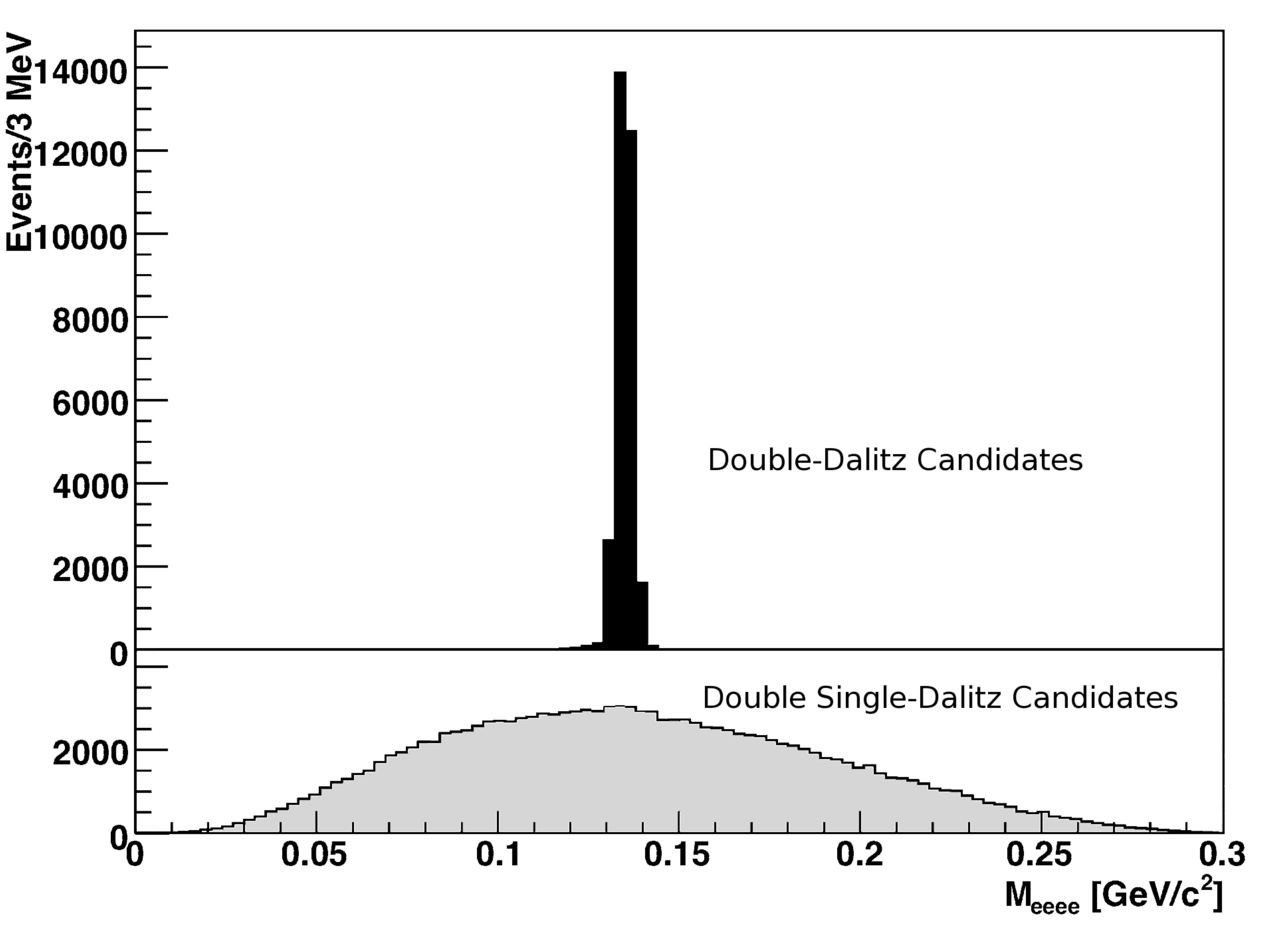}
\caption{\label{masspeak} Invariant $e^+e^-e^+e^-$ mass for data events
  identified by the preferred pairing as $\klpipipidd$
  (upper panel) and $\klpipidpid$ (lower panel). }
\end{figure}

The final selection criterion separates $\klpipipidd$ from
$\klpipidpid$ events. This is accomplished by a $\chi^2$ formed of the
three reconstructed $\pizero$ masses. This serves to identify the best
pairing of particles for a given decay hypothesis, as well as to
select the more likely hypothesis of the two. The event is tagged as
the mode with the smaller $\chi^2$, which is further required to be
less than $12$ (with three degrees of freedom).  This technique
correctly identifies more than $\units{99.5}{\%}$ of events
(Fig.~\ref{masspeak}).

The final event sample contains 30~511 signal candidates with
$\units{0.6}{\%}$ residual background and 141~251 normalization mode
candidates with $\units{0.5}{\%}$ background (determined from the
Monte Carlo simulation).  The background in the signal sample is
dominated by mistagged events from the normalization mode.

The branching ratio is measured from the ratio of reconstructed signal
mode events to normalization mode events.  This ratio must be
corrected by the ratio of acceptances, which has been determined using
a detailed Monte Carlo simulation of the beam distribution and
detector response.  The resulting double ratio is directly related to the
branching ratio $B_{eeee}$ of the $\pieeee$ mode normalized to the
square of the branching ratio $B_{ee\gamma}$ of the $\pieeg$ mode:
\begin{equation} \label{dblratio}
\frac{B_{eeee} \cdot B_{\gamma\gamma}}{B^{2}_{ee\gamma}}
 = \frac{N\left(\klpipipidd\right)}{N\left(\klpipidpid\right)} \cdot
   \frac{\epsilon\left(\klpipidpid\right)}{\epsilon\left(\klpipipidd\right)},
\end{equation}
where $N$ is the number of events and $\epsilon$ is the combined
geometric acceptance and detection efficiency for a given mode.  

The statistical error on the ratio in Eq.~\ref{dblratio} is 0.62\%.
Systematic errors on the efficiencies were determined through data
studies as well as variations in the parameters of the Monte Carlo
simulation.  Because the final state particles in the signal and
normalization mode are the same, detector-related quantities
substantially cancel in the ratio, which is generally insensitive to
the details of the simulation.  The dominant systematic errors came
from variation of the analysis cuts (0.21\%) and Monte Carlo
simulation statistics (0.25\%).  Other systematic errors
were from uncertainties in the amount of material
in the spectrometer (0.15\%), uncertainty in the background levels in
the two samples (0.15\%), modeling of the drift chamber
resolutions (0.11\%), and radiative corrections (0.04\%). 
The total systematic error on the relative
branching ratio is 0.41\%.

The final result for the ratio of decay rates is: 
\begin{equation}\begin{split} 
\frac{B_{eeee}^{ x>0.9} \cdot B_{\gamma\gamma}}{B^{2}_{ee\gamma}} = &
0.2245 \pm 0.0014{\rm (stat)} \pm 0.0009{\rm
  (syst)}.\end{split}\end{equation} The $\pieeee$ branching ratio can
be calculated from the double ratio using the known values
$B_{\gamma\gamma}= 0.9880\pm 0.0003$ and $B_{ee\gamma}= (1.198\pm
0.032)\times 10^{-2}$ \cite{dalitzbr}.  This yields $B_{eeee}^{x>0.9}=
(3.26 \pm 0.18)\times 10^{-5}$, where the error is dominated by the
uncertainty in the $\pieeg$ branching ratio.  Using our radiative
corrections model \cite{barker} to extrapolate to all radiative final states, we
find:
\begin{equation}\begin{split} 
\frac{B_{eeee(\gamma)} \cdot B_{\gamma\gamma}}{B^{2}_{ee\gamma}} = &
0.2383 \pm 0.0015{\rm (stat)} \pm 0.0010{\rm (syst)},\end{split}\end{equation}
and $B_{eeee(\gamma)}=(3.46\pm 0.19)\times 10^{-5}$.
Our branching ratio result is in agreement with
previous measurements \cite{samios2}.

The parameters of the $\piggvert$ coupling are found by maximizing
an unbinned likelihood function composed of the differential decay
rate in terms of ten phase-space variables.  The first five are $(x_1,
x_2, y_1, y_2, \phi)$, where $x_1$, $x_2$, and $\phi$ are described
above and the remaining variables $y_1$ and $y_2$ describe the energy
asymmetry between the electrons in each Dalitz pair in the $\pi^0$
center of mass \cite{barker}.  The remaining five are the same variables, but
calculated with the opposite choice of $e^+e^-$ pairings.  The
likelihood is calculated from the full matrix element including the
exchange diagrams and $\mathcal{O}(\alpha^2)$ radiative corrections.

The fit yields the DIP $\alpha$ parameter and the (complex) ratio of
the scalar to the pseudoscalar coupling.  For reasons of fit
performance, the parity properties are fit to the equivalent
parameters $\kappa$ and $\eta$, where $\kappa + \mi \eta \equiv \tan
\zeta \me^{\mi\delta}$. The shape of the minimum of the likelihood
function indicates that the three parameters $\alpha$, $\kappa$, and
$\eta$ are uncorrelated.  Acceptance-dependent effects are included as
a normalization factor calculated from Monte Carlo simulations.

Systematic error sources on $\alpha$ and $\kappa$ are similar to those
for the branching ratio measurement.  The dominant error is due to
variation of cuts, resulting in a total systematic error of 0.9 and
0.011 on $\alpha$ and $\kappa$ respectively. For the $\eta$ parameter,
the primary uncertainty results from the resolution on the angle
$\phi$ between the two lepton pairs, which produces an effective
flattening of the angular distribution without inducing a phase shift.
The fitter interprets this as a small scalar contribution with a phase
difference of $90$~degrees, and therefore a larger value of $\eta$,
particularly for $\eta\approx 0$. This behavior was studied
with Monte Carlo simulation and a correction was calculated. The
uncertainty on this correction results in a systematic error of 0.031.

\begin{figure}
\includegraphics[width=3.4in]{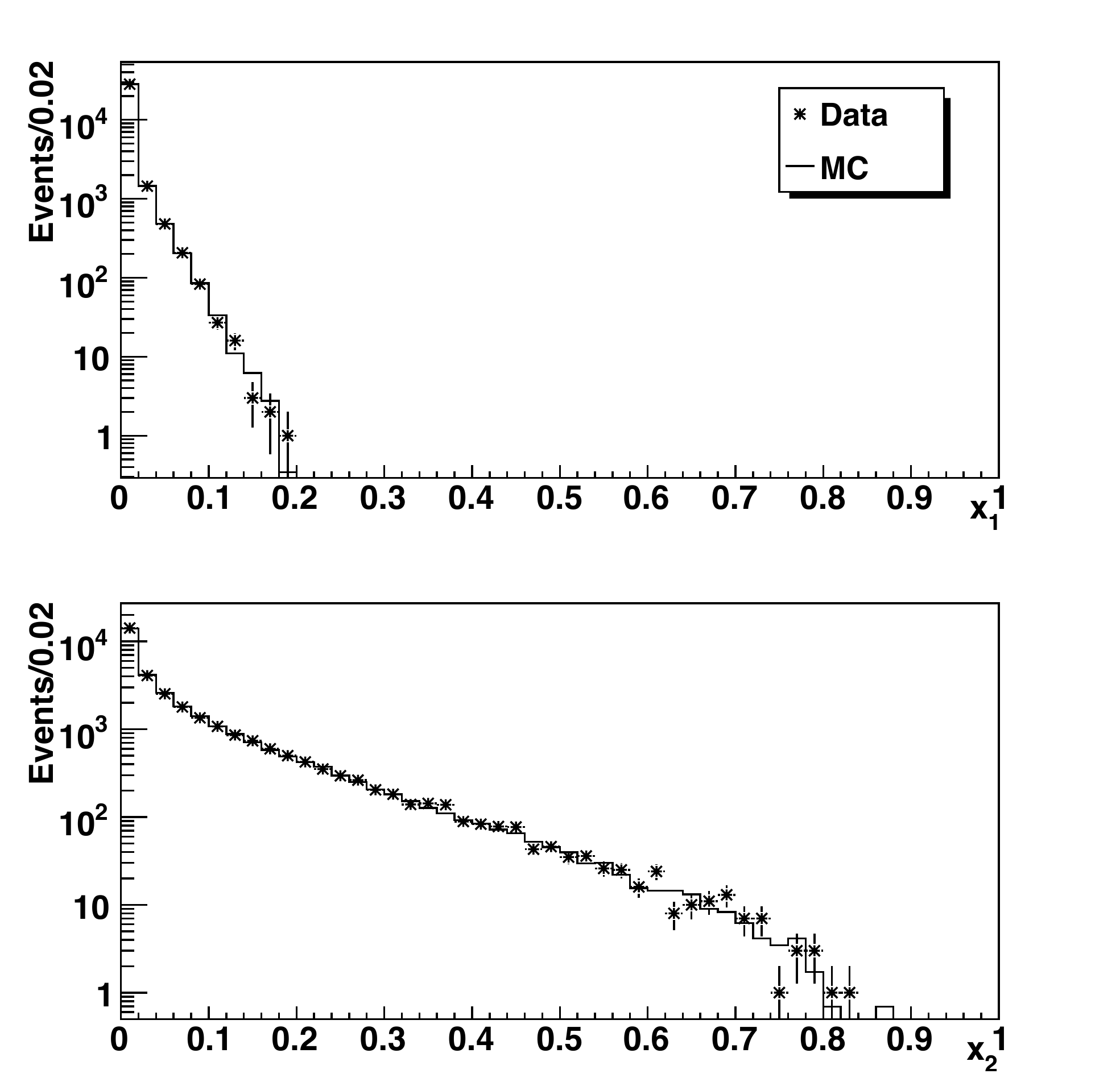}
\caption{\label{x1x2} Distribution of the kinematic variables $x_1$ and
$x_2$ for signal event candidates (points) and signal Monte Carlo simulation
with best fit form factor parameters (histogram).}
\end{figure}

\begin{figure}
\includegraphics[width=3.4in]{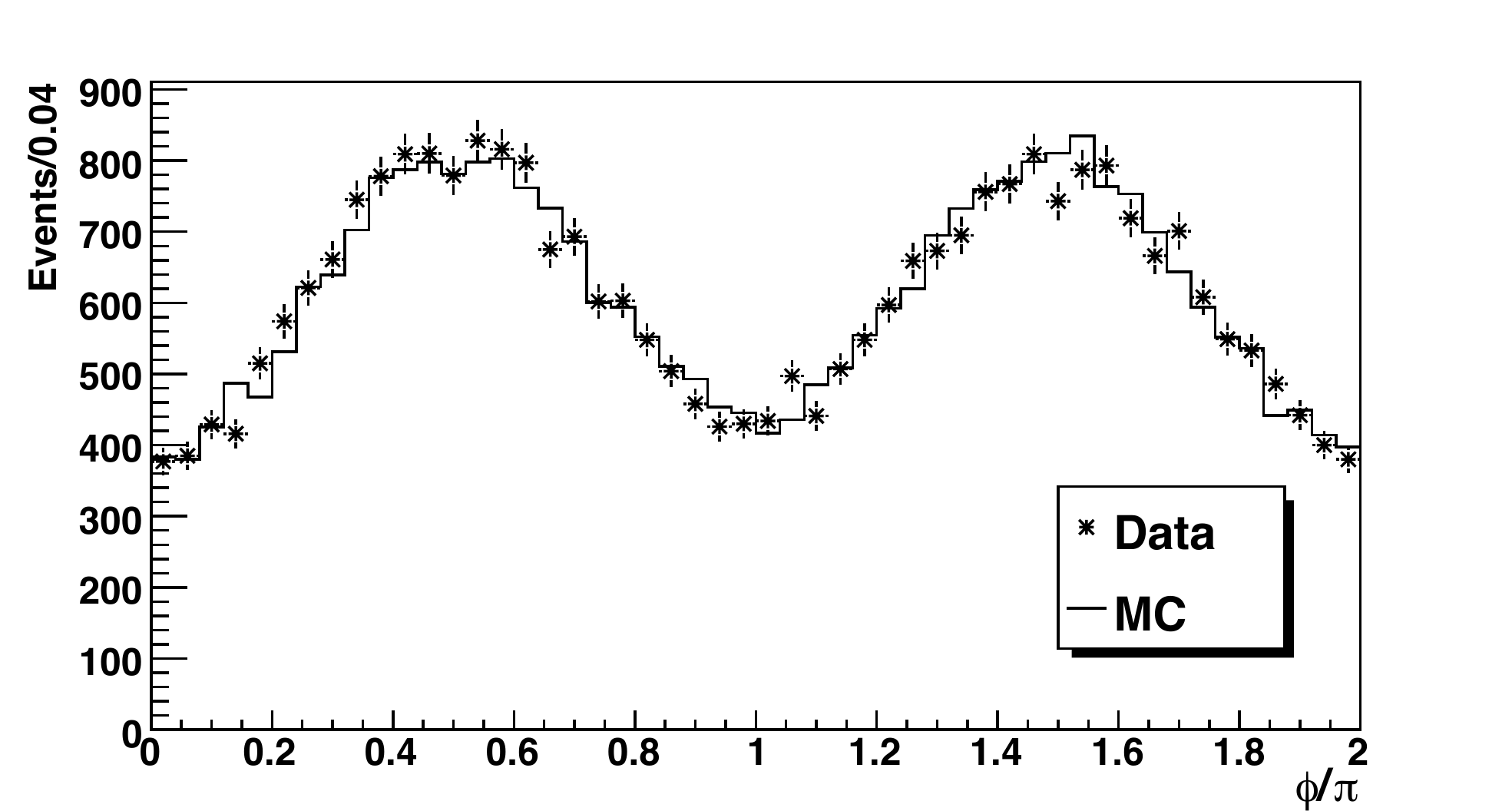}
\caption{\label{phi} Distribution of the angle $\phi$, in units of
  $\pi$, between the planes of the two $e^+e^-$ pairs.  The solid
  histogram shows the Monte Carlo expectation for negative parity.}
\end{figure}

The distributions of $x_1$ and $x_2$, overlaid with the Monte Carlo simulation,
are shown in Fig.\ \ref{x1x2}.  The
$\phi$ distribution is shown in Fig.\ \ref{phi}.  For plotting the
data a unique pairing of the four electrons is chosen such that
$x_1<x_2$ and the product $x_1x_2$ is minimized: this choice
represents the dominant contribution to the matrix element.  It is
clear that the pseudoscalar coupling dominates, as expected, with no
evidence for a scalar component.  The distributions of all
five phase space variables agree well with the Monte Carlo simulation.

The final results for the three parameters are $\alpha = 1.3 \pm 1.0 {\rm (stat)} \pm 0.9 {\rm (syst)}$,
$\kappa = -0.011 \pm 0.009 {\rm (stat)} \pm 0.011 {\rm (syst)}$, and $\eta = 0.051 \pm 0.026 {\rm (stat)} \pm 0.031 {\rm (syst)}$.  The DIP
$\alpha$ parameter is related to the standard slope parameter by $a =
-0.032 \alpha$, yielding $a = -0.040 \pm 0.040$. 
This result is in agreement with recent direct measurements. 

The parameters $\kappa$ and $\eta$
are transformed into limits on the pseudoscalar-scalar mixing angle
$\zeta$ under two hypotheses. If $\oCPT$ violation is allowed, then
the limit is set by the uncertainties in $\eta$, resulting in $\zeta <
6.9^\circ$ at the $90\%$ confidence level. If instead, $\oCPT$
conservation is enforced, $\eta$ must be zero, and the limit derives
from the uncertainties on $\kappa$, resulting in $\zeta < 1.9^\circ$,
at the same confidence level. These limits on $\zeta$ limit the
magnitude of the scalar component of the decay amplitude, relative to the
pseudoscalar component, to less than $12.1\%$ in the presence of $\oCPT$
violation, and less than $3.3\%$ if $\oCPT$ is assumed conserved.  The
limits on scalar contributions apply to all $\pi^0$ decays with 
two-photon intermediate or final states.

This analysis confirms the negative parity of the neutral pion with
much higher statistical significance than the previous result, and
places tight limits on nonstandard scalar and $\oCPT$-violating
contributions to the $\pieeee$ decay.  We have also measured the
momentum dependent form factor in this decay for the first time, and
made the first improvement in its branching ratio since 1962.  This
measurement is limited at present by the current large uncertainty in
the branching ratio of the single Dalitz decays used for normalization, but we
expect that uncertainty to be reduced in the near future at which
point the present measurement can be recalculated using the more
precise double ratio measurement.

We gratefully acknowledge the support and effort of the Fermilab
staff and the technical staffs of the participating institutions for
their vital contributions.  This work was supported in part by the
U.S.\ Department of Energy, the National Science Foundation,
the Ministry of Education and Science of Japan, Fundao de Amparo a
Pesquisa do Estado de S Paulo-FAPESP, Conselho Nacional de Desenvolvimento
Cientifico e Tecnologico-CNPq and CAPES-Ministerio Educao.

\begin{thebibliography}{13}
\expandafter\ifx\csname natexlab\endcsname\relax\def\natexlab#1{#1}\fi
\expandafter\ifx\csname bibnamefont\endcsname\relax
  \def\bibnamefont#1{#1}\fi
\expandafter\ifx\csname bibfnamefont\endcsname\relax
  \def\bibfnamefont#1{#1}\fi
\expandafter\ifx\csname citenamefont\endcsname\relax
  \def\citenamefont#1{#1}\fi
\expandafter\ifx\csname url\endcsname\relax
  \def\url#1{\texttt{#1}}\fi
\expandafter\ifx\csname urlprefix\endcsname\relax\def\urlprefix{URL }\fi
\providecommand{\bibinfo}[2]{#2}
\providecommand{\eprint}[2][]{\url{#2}}

\bibitem[{\citenamefont{Panofsky et~al.}(1951)\citenamefont{Panofsky, Aamodt,
  and Hadley}}]{panofsky}
\bibinfo{author}{\bibfnamefont{W.~K.~H.} \bibnamefont{Panofsky}},
  \bibinfo{author}{\bibfnamefont{R.~L.} \bibnamefont{Aamodt}},
  \bibnamefont{and} \bibinfo{author}{\bibfnamefont{J.}~\bibnamefont{Hadley}},
  \bibinfo{journal}{Phys. Rev.} \textbf{\bibinfo{volume}{81}},
  \bibinfo{pages}{565} (\bibinfo{year}{1951}).

\bibitem[{\citenamefont{Chinowsky and Steinberger}(1951)}]{chinowsky}
\bibinfo{author}{\bibfnamefont{W.}~\bibnamefont{Chinowsky}} \bibnamefont{and}
  \bibinfo{author}{\bibfnamefont{J.}~\bibnamefont{Steinberger}},
  \bibinfo{journal}{Phys. Rev.} \textbf{\bibinfo{volume}{95}},
  \bibinfo{pages}{1561} (\bibinfo{year}{1951}).

\bibitem[{\citenamefont{Yang}(1950)}]{yang}
\bibinfo{author}{\bibfnamefont{C.~N.} \bibnamefont{Yang}},
  \bibinfo{journal}{Phys. Rev.} \textbf{\bibinfo{volume}{77}},
  \bibinfo{pages}{242} (\bibinfo{year}{1950}).

\bibitem[{\citenamefont{Bernstein and Michel}(1950)}]{bernstein}
\bibinfo{author}{\bibfnamefont{J.}~\bibnamefont{Bernstein}} \bibnamefont{and}
  \bibinfo{author}{\bibfnamefont{L.}~\bibnamefont{Michel}},
  \bibinfo{journal}{Phys. Rev.} \textbf{\bibinfo{volume}{118}},
  \bibinfo{pages}{871} (\bibinfo{year}{1950}).

\bibitem[{\citenamefont{Kroll and Wada}(1955)}]{kroll}
\bibinfo{author}{\bibfnamefont{N.~M.} \bibnamefont{Kroll}} \bibnamefont{and}
  \bibinfo{author}{\bibfnamefont{W.}~\bibnamefont{Wada}},
  \bibinfo{journal}{Phys. Rev.} \textbf{\bibinfo{volume}{98}},
  \bibinfo{pages}{1355} (\bibinfo{year}{1955}).

\bibitem[{\citenamefont{Dalitz}(1951)}]{dalitz}
\bibinfo{author}{\bibfnamefont{R.~H.} \bibnamefont{Dalitz}},
  \bibinfo{journal}{Proc. Phys. Soc. (London)} \textbf{\bibinfo{volume}{A64}},
  \bibinfo{pages}{667} (\bibinfo{year}{1951}).

\bibitem[{\citenamefont{Samios et~al.}(1962)\citenamefont{Samios, Plano,
  Prodell, Schwartz, and Steinberger}}]{samios2}
\bibinfo{author}{\bibfnamefont{N.~P.} \bibnamefont{Samios}},
  \bibinfo{author}{\bibfnamefont{R.}~\bibnamefont{Plano}},
  \bibinfo{author}{\bibfnamefont{A.}~\bibnamefont{Prodell}},
  \bibinfo{author}{\bibfnamefont{M.}~\bibnamefont{Schwartz}}, \bibnamefont{and}
  \bibinfo{author}{\bibfnamefont{J.}~\bibnamefont{Steinberger}},
  \bibinfo{journal}{Phys. Rev.} \textbf{\bibinfo{volume}{126}},
  \bibinfo{pages}{1844} (\bibinfo{year}{1962}).

\bibitem[{\citenamefont{Barker et~al.}(2003)\citenamefont{Barker, Huang, Toale,
  and Engle}}]{barker}
\bibinfo{author}{\bibfnamefont{A.~R.} \bibnamefont{Barker}},
  \bibinfo{author}{\bibfnamefont{H.}~\bibnamefont{Huang}},
  \bibinfo{author}{\bibfnamefont{P.~A.} \bibnamefont{Toale}}, \bibnamefont{and}
  \bibinfo{author}{\bibfnamefont{J.}~\bibnamefont{Engle}},
  \bibinfo{journal}{Phys. Rev. D} \textbf{\bibinfo{volume}{67}},
  \bibinfo{pages}{033008} (\bibinfo{year}{2003}).

\bibitem[{\citenamefont{Fonvieille et~al.}(1989)}]{fonvieille}
\bibinfo{author}{\bibfnamefont{H.}~\bibnamefont{Fonvieille}}
  \bibnamefont{et~al.}, \bibinfo{journal}{Phys. Lett. B}
  \textbf{\bibinfo{volume}{233}}, \bibinfo{pages}{65} (\bibinfo{year}{1989}).

\bibitem[{\citenamefont{Farzanpay et~al.}(1992)}]{farzanpay}
\bibinfo{author}{\bibfnamefont{F.}~\bibnamefont{Farzanpay}}
  \bibnamefont{et~al.}, \bibinfo{journal}{Phys. Lett. B}
  \textbf{\bibinfo{volume}{278}}, \bibinfo{pages}{413} (\bibinfo{year}{1992}).

\bibitem[{\citenamefont{Drees et~al.}(1992)}]{drees}
\bibinfo{author}{\bibfnamefont{R.~M.} \bibnamefont{Drees}}
  \bibnamefont{et~al.}, \bibinfo{journal}{Phys. Rev. D}
  \textbf{\bibinfo{volume}{45}}, \bibinfo{pages}{1439} (\bibinfo{year}{1992}).

\bibitem[{\citenamefont{D'Ambrosio et~al.}(1998)\citenamefont{D'Ambrosio,
  Isidori, and Portol\'{e}s}}]{dip}
\bibinfo{author}{\bibfnamefont{G.}~\bibnamefont{D'Ambrosio}},
  \bibinfo{author}{\bibfnamefont{G.}~\bibnamefont{Isidori}}, \bibnamefont{and}
  \bibinfo{author}{\bibfnamefont{J.}~\bibnamefont{Portol\'{e}s}},
  \bibinfo{journal}{Phys. Lett. B} \textbf{\bibinfo{volume}{423}},
  \bibinfo{pages}{385} (\bibinfo{year}{1998}).

\bibitem[{\citenamefont{P.\ A.\ Toale}(2004)\citenamefont{P.\ A.\ Toale}}]{toalethesis} 
\bibinfo{author}{\bibfnamefont{P.}~\bibfnamefont{A.}~\bibnamefont{Toale}},
  \bibinfo{journal}{Ph.D.\ dissertation, The University of Colorado}
  (\bibinfo{year}{2004}).

\bibitem[{\citenamefont{Abouzaid et~al.}(2007)}]{ktev799}
\bibinfo{author}{\bibfnamefont{E.}~\bibnamefont{Abouzaid}}
  \bibnamefont{et~al.} (\bibinfo{collaboration}{KTeV}), \bibinfo{journal}{Phys.
  Rev. D} \textbf{\bibinfo{volume}{75}}, \bibinfo{pages}{012004}
  (\bibinfo{year}{2007}).

\bibitem[{\citenamefont{Alavi-Harati et~al.}(2003)}]{ktev_eps}
\bibinfo{author}{\bibfnamefont{A.}~\bibnamefont{Alavi-Harati}}
  \bibnamefont{et~al.} (\bibinfo{collaboration}{KTeV}), \bibinfo{journal}{Phys.
  Rev. D} \textbf{\bibinfo{volume}{67}}, \bibinfo{pages}{012005}
  (\bibinfo{year}{2003}).

\bibitem[{\citenamefont{M. A. Schardt et~al.}(2003)}]{schardt}
\bibinfo{author}{\bibfnamefont{M.~A.}~\bibnamefont{Schardt}}
  \bibnamefont{et~al.}, \bibinfo{journal}{Phys.
  Rev. D} \textbf{\bibinfo{volume}{23}}, \bibinfo{pages}{639}
  (\bibinfo{year}{1981}).

\bibitem[{\citenamefont{PDG}(2006)}]{dalitzbr}
\bibinfo{author}{\bibfnamefont{W.-M.}~\bibnamefont{Yao}}
  \bibnamefont{et~al.}, \bibinfo{journal}{J.~Phys.~G}
  \textbf{\bibinfo{volume}{33}}, \bibinfo{pages}{1}
  (\bibinfo{year}{2006}).

\end{thebibliography}

\end{document}